\documentclass[runningheads]{llncs}
\usepackage[T1]{fontenc}
\usepackage{graphicx}
\usepackage{url,hyperref,lineno,microtype}
\usepackage[onehalfspacing]{setspace}
\usepackage{physics}
\usepackage{caption}
\usepackage{float}
\usepackage{algorithm}
\usepackage{algpseudocode}
\usepackage{amsmath}
\usepackage{multirow}
\usepackage{amsfonts}

\begin{document}
\title{Biclustering a dataset using photonic quantum computing}
\titlerunning{Biclustering with photonic quantum computing}
%
\author{Ajinkya Borle\inst{1} \and
Ameya Bhave\inst{1} }
\authorrunning{Borle et al.}
%
\institute{CSEE Department, University of Maryland Baltimore County, Baltimore, MD 21250\\
\email{\{aborle1,ameyab1\}@umbc.edu}}
\maketitle              
\begin{abstract}
Biclustering is a problem in machine learning and data mining that seeks to group together rows and columns of a dataset according to certain criteria. In this work, we highlight the natural relation that quantum computing models like boson and Gaussian boson sampling (GBS) have to this problem. We first explore the use of boson sampling to identify biclusters based on matrix permanents. We then propose a heuristic that finds clusters in a dataset using Gaussian boson sampling by (i) converting the dataset into a bipartite graph and then (ii) running GBS to find the densest sub-graph(s) within the larger bipartite graph. Our simulations for the above proposed heuristics show promising results for future exploration in this area.

\end{abstract}
\section{Introduction}

Quantum machine learning is an emerging field of study that is at the intersection of quantum physics and machine learning. It contains research problems that span from leveraging quantum computing for machine learning, to the use of machine learning methods to model quantum physics. Broadly speaking, our work is in the former set of problems. We study the use of computational models enabled by photonics, specifically boson sampling and Gaussian boson sampling (GBS), for an unsupervised learning problem: biclustering.

Biclustering or Co-clustering is the selection of rows and columns of a matrix based on a given criteria (largest values, similar values, constant values, etc) \cite{mirkin1997mathematical}. It currently has applications in (but not limited to) bioinformatics \cite{pontes2015biclustering,xie2019time,castanho2022biclustering},text mining \cite{de2007applying,orzechowski2016text}, recommender systems \cite{choi2018reinforcement,sun2022recommendation}, and even fields like malware analysis \cite{raff2020automatic}. While the time complexity of biclustering depends on the exact formulation of the problem (i.e. the criteria for the biclusters), the problems of particular interest are the ones for which the decision problems that are NP-complete in nature; e.g. does this matrix have a bicluster of size $b_{1} \times b_{2}$? An answer to such a question can be verified in polynomial time, but not found in polynomial time \cite{cormen2022introduction}. Therefore, meta-heuristics are often used for this task \cite{jose2022biclustering}, to get good solutions fast, but without theoretical guarantees.

Among the many models of quantum computing, certain metaheuristic based models have also been proposed on which relevant machine learning and data mining problems can be mapped onto \cite{adachi2015application,kumar2018quantum,schuld2020measuring,bonaldi2023boost}. These include quantum annealing \cite{kadowaki1998quantum}, Boson sampling \cite{aaronson2011computational} and GBS \cite{hamilton2017gaussian}. While a method to apply quantum annealing to the task of biclustering already exists \cite{bottarelli2018biclustering}, our work, to the best of our knowledge, is the first one that applies boson sampling and GBS to this problem.

Boson Sampling is a restricted model of quantum computing most easily implemented with photonic quantum computing, in particular, with linear optics \cite{aaronson2011computational}. GBS is a variant of the above that generates photons by squeezing light \cite{hamilton2017gaussian}. Both models solve \#P-hard problems \footnote{A complexity class which has problems that are at least as hard as the hardest problems in NP.} with proposed applications in the fields of graph theory \cite{mezher2023solving}, machine learning\cite{schuld2020measuring,bonaldi2023boost}, and optimization \cite{arrazola2018quantum} among others. These are quantum computing models that when applied with photonic quantum computing, are feasible right in the NISQ era of quantum computing but can have utility even beyond it \cite{madsen2022quantum,deshpande2022quantum}.

The contributions of this paper are as follows:
\begin{enumerate}
    \item We propose and explore the application of boson sampling (and Gaussian boson sampling) to the problem of biclustering in machine learning. To the best of our knowledge, this is the first work to do so.
    \item For boson sampling, we applied the unitary dilation theorem \cite{halmos1950summa,mezher2023solving} to embed our dataset in an unitary matrix.
    \item We propose a simulated annealing (SA) technique that uses boson sampling as a subroutine for finding biclusters.
    \item For GBS, we show how to embed a dataset as a unitary matrix. This is done by first considering the dataset as a bipartite graph \cite{karim2019bicluso} and then using the Autonne-Takagi decomposition \cite{takagi1924algebraic} on it.
    \item We performed preliminary simulations to study the basic characteristics of both boson sampling and GBS for the task of detecting biclusters in a dataset.
\end{enumerate}

Since our work is the first one for this topic, our focus was on establishing the basics that would be crucial for any follow-up research done in the field. Our results show conditions for when boson sampling and GBS do well (e.g. for binary datasets) and certain situations to watch out for (some problems may need lot more samples than others). We believe that our results can give useful insights for future work in this direction. 

\section{Background}\label{sec:background}
In this section, we will cover the topics and notations that will be used in this work.
\subsection{Permanent}\label{sec:perm}
A Permanent of an $N \times N$ matrix $A$ is an operation that can mathematically be defined as: 
\begin{align}
    \text{Per}(A) = \sum_{\sigma \in S_N}\prod_{i=1}^{N}A_{i,\sigma(i)}
\end{align}

Where $\sigma$ is a permutation of the symmetric group $S_N$. In other words, a permanent is the summation of the products of all the possible elements with all permutations of index values for rows and columns \footnote{In any given product, no row or column can be repeated.}. This is similar to calculating the determinant using the Leibinz's rule \cite{miller1930history} except that the sign of each summand is positive.

Calculating the permanent is a \#P-hard problem that takes exponential time even for the best classical algorithms \cite{ryser1963combinatorial,glynn2013permanent}. It is an important component in the analysis of the probabilities from boson sampling.
\subsection{Boson sampling}\label{sec:boson_sampling}
Boson sampling is a non-universal model of quantum computation, pioneered by Aaronson and Arkhipov in 2011 \cite{aaronson2011computational} on the observations of Troyansky et al. \cite{troyansky1996quantum}. A boson is a subatomic particle that has an integer spin number (e.g. Higgs boson, photon, gluon, etc). More importantly for us, the most popular and feasible approach for its realization is by the use of photons. For the remainder of the paper, we would be using terminology from quantum optics for this process.

Unlike other models of quantum computing, instead of using qubits as the building blocks of quantum information, we use linear optical modes (through which photons can traverse) for carrying and manipulating information. Here, we assume $m$ modes carrying $n$ photons (where $m > n$), are plugged into a linear-optical network made up of beamsplitters and phaseshifters. The linear-optical network can be represented as a Unitary matrix $U \in \mathbb{C}^{m \times m}$. Here we define the input state as
\begin{align}
    \ket{\psi_{0}} &= \ket{n_1,n_2,n_3,...,n_m} \label{eq:input_fock} \\
    \text{such that } n &= \sum_{i=1}^{m}n_i \\
    \text{and } n_i &\geq 0
\end{align}

Where Eqn(\ref{eq:input_fock}) is a Fock state that denotes the number of photons in the input state for each of the modes involved. After passing through the linear-optical network, we measure the number of photons in each mode. Let us denote this state\footnote{Assuming there is no photon loss.} by $\ket{\psi'}$
\begin{align}
    \ket{\psi'} &= \ket{n_{1}',n_{2}',n_{3}',...,n_{m}'} \label{eq:output_fock} \\
    \text{such that } n &= \sum_{i=1}^{m}n_{i}' \label{eq:sum_observed_photons} \\
    \text{and } n_i' &\geq 0
\end{align}
Meaning that the number of photons in each mode may have changed at the end of the computation. The probability of a particular Fock state to be measured at the end of a single run (or sample) of this process (given an input state) is denoted by
\begin{align}
    P(\psi'|\psi_{0}) = \frac{|\text{Per}(U_{\psi',\psi_0})|^2}{n_{1}!,n_{2}!,n_{3}!,...,n_{m}!,n_{1}'!,n_{2}'!,n_{3}'!,...,n_{m}'!} \label{eq:probability_bs}
\end{align}
Where $U_{\psi',\psi_0}$ is an $n \times n$ sub-matrix constructed from $U$ by taking $n_{i}$ times
the ith column on U, and $n_{j}'$ times the jth row of $U$, $\forall$ $1 \leq i,j \leq m$\cite{aaronson2011computational,mezher2023solving}.

Sub-matrices that have larger permanents will have a higher probability of being measured. While for a mode, any number of photons larger than 1 will have an adverse effect on the probability. With enough sampling, we get the probability distribution described by Eqn(\ref{eq:probability_bs}). Typically for boson sampling, it is often assumed that our starting state $\ket{\psi_0}$ has 1 photon at maximum per mode. And for a variety of problems, it is also expected that the states we are interested in have 1 photon at maximum for a mode in the output \cite{aaronson2011computational}. Thus, in order for us to apply boson sampling to real world problems, we need to encode them as unitary matrices where the solutions can be decoded from sub-matrices with the largest permanent values.

\subsection{Hafnian}\label{sec:haf}
Related to the permanent, the Hafnian \cite{Termini} of a symmetric matrix $A \in \mathbb{C}^{N \times N}$ is a value that is calculated by the following equation
\begin{align}
    \text{Haf}(A) = \sum_{\rho \in P}\prod_{\{i,j\}\in\rho}A_{i,j}
\end{align}

Where $P$ is the set of perfect matchings for a fully-connected graph of $N$ vertices\footnote{This does not mean that $A$ is fully-connected, it just means that the number of matchings considered for the calculation of the Hafnian are from a fully connected graph (that has the same number of vertices as $A$ does).}. The permanent of a matrix $C$ and its hafnian are connected by the following relationship
\begin{align}
    \text{Per}(C) = \text{Haf}\big( \begin{pmatrix}
    0 & C\\
    C^T & 0
    \end{pmatrix} \big)
\end{align}
\subsection{Gaussian boson sampling (GBS)}\label{sec:gbs}
One of the biggest challenges in the implementation of boson sampling is the production of synchronized single-photons on a large scale. Different schemes of producing photons have been suggested in order to address this, such as Gaussian boson sampling \cite{hamilton2017gaussian}.

In the physical setup for GBS (implemented using photonics), the linear interferometer for $m$ modes is prefixed with squeezing operators on all the modes individually. This does not produce an exact number of photons but depending on the squeezing parameter, can produce photons with an average count per mode from a Gaussian distribution\footnote{To be considered as an hyperparameter.} (henceforth known as mean number of photons $\overline{n}$ per mode) . The unitary matrix that defines the linear interferometer is typically constructed from a symmetric matrix $A \in \mathbb{C}^{N \times N}$ (where $m = N$)\footnote{Not considering any hardware restrictions.} after using a process known as Autonne-Takagi decomposition \cite{takagi1924algebraic}. 

At the end of the computation, the number of photons that appear at each mode are read out $\ket{\psi'} = \ket{n_{1}'n_{2}'...n_{m}'}$. The probability of reading a particular $\ket{\psi'}$ is proportional to
\begin{align}
    P(\psi') \propto c^n\frac{\text{Haf}(A_{\psi'})}{n_{1}'!n_{2}'!...n_{m}'!} \label{eq:probability_gbs}
\end{align}
Where $c$ is a scaling parameter from squeezing and $n$ is the sum of all observed photons (Eqn(\ref{eq:sum_observed_photons}). The matrix $A_{\psi'}$ is an $n \times n$ matrix constructed by taking $n_i$ times the ith column and ith row of $A$, for $1\leq i \leq m$. For many computational problems however, as a simplified heuristic, the submatrix at the end is constructed from the unique rows and columns indicated by the ith mode for which $n_{i} > 0$. This is typically done with threshold detectors for measurement and the exact probability distribution for such a setup depends on a matrix function called torontonian \cite{quesada2018gaussian,deng2023solving} (a function that is analogous to hafnian). The task then becomes to encode real world problems for which the submatrix with a high hafnian (or torontonian to be more specific) value would yield us the solution.

Readers who want to learn about GBS in depth are recommended to read the original paper \cite{hamilton2017gaussian}.

\subsection{The biclustering problem} \label{sec:background_bicluster}
Like previously mentioned, the problem of biclustering is one where the rows and columns of a matrix are clustered together (called biclusters) depending on a criteria. Biclusters can be formed by different criteria \cite{hochreiter2010fabia}, some of the popular ones are :
\begin{enumerate}
    \item Biclusters with a constant value populating all the cells, constant value for each row or constant value for each column.
    \item Biclusters where values are unusually high or low (with respect to the rest of the matrix).
    \item Biclusters that have low variance.
    \item Biclusters that have correlated rows and columns.
\end{enumerate}
For our work, we will focus on biclusters for two criteria (i) biclusters that have high values and (ii) biclusters that have maximum number of ones in a binary matrix\footnote{Assuming all values lie in the range $[0,1]$, we can also potentially find (iii) biclusters with low values and (iv) biclusters with majority zeros for a binary matrix. We can do this by subtracting current value of each cell in the matrix from 1.}. All matrices (representing datasets) and sub-matrices (representing biclusters) considered in our simulations are square shaped, but we also comment on how our work can be extended for rectangular datasets and biclusters (in the following sections).

\subsection{Simulated Annealing (SA)}

Simulated annealing (SA) is a family of heuristics that aim to optimize a cost function by randomly sampling from the solution space of an objective function and then accepting or rejecting a new sample based on the current sample and a temperature parameter \cite{bertsimas1993simulated}. In the context of our work, we use it as a black box optimization method with boson sampling as a subroutine for finding the best columns for a bicluster. Section \ref{sec:bs_exp1} and algorithm \ref{alg:bs_bicluster} (lines \ref{line:SA_1}, \ref{line:SA_2} and \ref{line:SA_3} in particular) describe how SA is implemented in our work. For the readers who want to know more about the fundamentals of the original technique, we recommend the original paper by Kirkpatrick et. al. \cite{kirkpatrick1983optimization}.
\begin{figure}[t]
  \centering
   \includegraphics[scale=0.6]{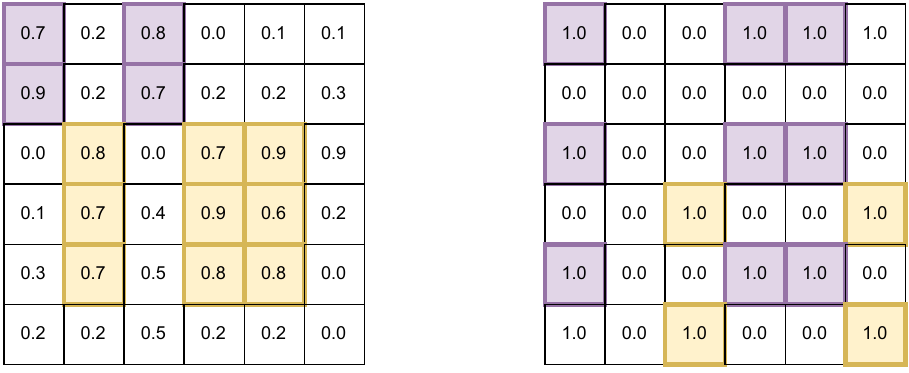}  
\caption{Examples of square biclusters in a larger square matrix, representing a dataset. (LEFT) is an example of a dataset that has elements in the range $[0,1]$ and (RIGHT) is an example of a dataset that has binary elements. In each dataset, there exist two biclusters that are distingushable by different colors.}
\label{fig:bicluster_example}
\end{figure}
\section{Related work}\label{sec:rw}
The field of biclustering has been well studied from the perspective of classical computing. A lot of progress that has been made in creating or improving biclustering algorithms has come from the field of bioinformatics \cite{madeira2004biclustering,ayadi2009biclustering,castanho2022biclustering}. While the most well-known algorithms for finding biclustering are based on spectral decomposition of matrices \cite{dhillon2001co,kluger2003spectral}, plenty of other heuristics have also been explored for this task \cite{hochreiter2010fabia,jose2022biclustering}.

Previously, the application of quantum annealing to the problem of biclustering was proposed \cite{bottarelli2018biclustering} where the problem of finding biclusters was encoded in the quadratic unconstrained binary optimization (QUBO) form and then solved on a D-wave\textsuperscript{TM} 2X annealer\footnote{Technically, QUBO problems are first converted to their Ising model \cite{cipra1987introduction} equivalents and then run on a quantum annealer.}. The authors performed experiments on datasets of size $100 \times 50$ but with a focus on a $10 \times 10$ moving window with biclusters upto $6 \times 6$ in size.

As far as photonic quantum computing is concerned, our work is the first that attempts to apply boson sampling and GBS to this problem. At the time of writing this paper, photonic quantum computing is still in its nascent stage\footnote{But with the potential of scaling up in a noise-resistant manner.}, and our primary aim is to gain preliminary insights about the relationship between photonic quantum computing and biclustering. For this, we worked with dataset of size $12 \times 12$ and biclusters of sizes $4\times 4$ and $6\times 6$.

The other work that is related to ours is the work done on using GBS for clustering \cite{bonaldi2023boost}, where the authors showed promising results (on simulators) when compared to results produced by classical methods. In their work, Bonaldi et. al. created a graph for GBS whose edges represented inverse distances between datapoints (vertices) and found clusters by setting the weight of the edge to 0 or 1 according to a threshold. In our work, we do not use inverse distances as a concept and have a different method for encoding the datasets into linear interferometers.

\section{Boson sampling for biclustering}
\subsection{Introduction}
Let us consider a dataset represented by a matrix $D \in \mathbb{R}^{d_{1} \times d_{2}}$ in which we need to find $k$ biclusters $\mathcal{B} =\{\beta_{1},\beta_{2},...,\beta_{k}\}$ of size\footnote{We will mention how to embed rectangular matrices and search for rectangular biclusters later in this work.} $b \times b$ each. Like mentioned in section \ref{sec:background_bicluster}, we are looking for biclusters with large values. Our method is based on the conjecture that large values in a bicluster would imply that the bicluster has a large permanent.

We use boson sampling as a subroutine to find out which $b$ rows would give the highest permanent value for a given set of $b$ columns. We do this over many different sets of columns to finally end up with a bicluster matrix that has the largest permanent. We can then set all the values of that bicluster to 0 (in the dataset) and repeat the process again for new biclusters (up until a termination criteria is met\footnote{The termination criteria can be anything: from a maximum number of biclusters to a threshold value of the bicluster's norm.}). 

\begin{figure}
  \centering
   \includegraphics[scale=0.53]{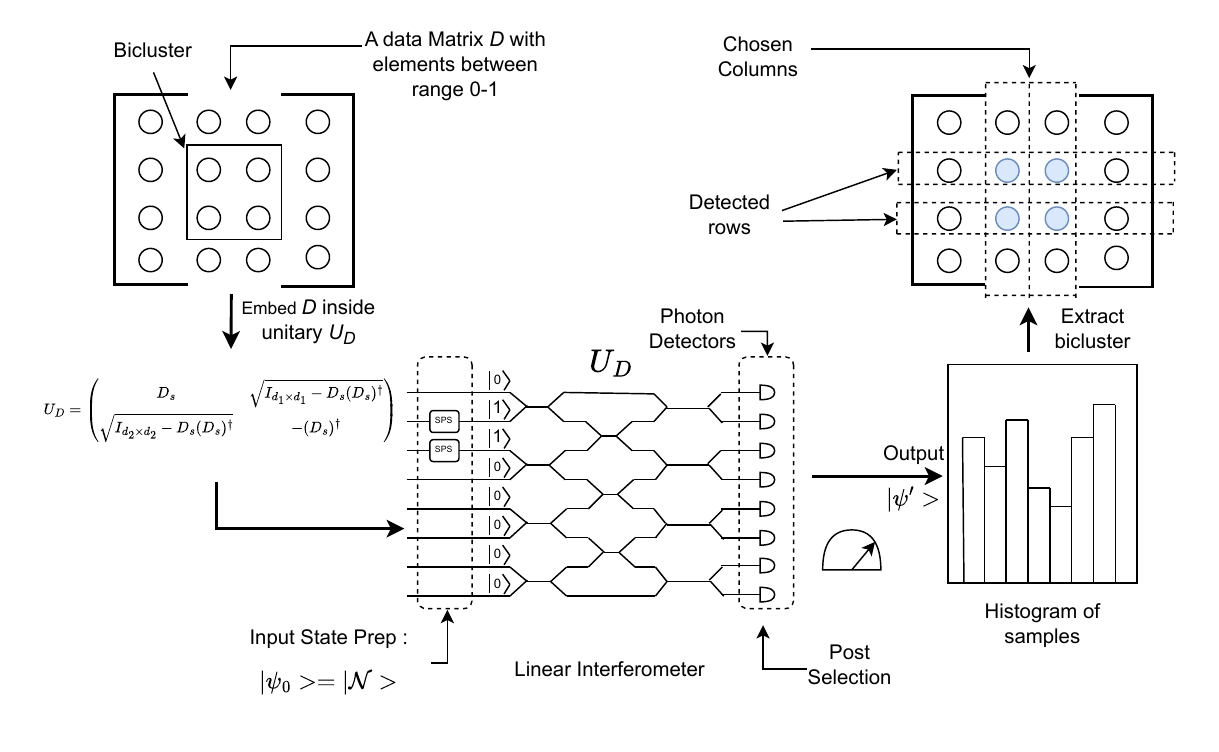}  
\caption{Workflow of the boson sampling approach for biclustering. Here, the user chooses the columns and boson sampling returns corresponding rows; from which the candidate bicluster is constructed. By Eqn(\ref{eq:probability_bs}), it is expected that the bicluster that has the highest permanent value would also have the highest probability of being obtained (for the initial choice of columns). This figure is for illustrative purposes only.}
\label{fig:boson_sampling_flowchart}
\end{figure}
\subsection{Approach}
In order to use boson sampling, we first need to embed our matrix $D$ as a unitary operator. For simplification, we assume that all of the elements are real values scaled to the interval of $[0,1]$. We first scale this matrix by $s=\sigma_{max}(D)$, the largest singular value of $D$ to produce $D_{s}= \frac{1}{s} D$. We then embed this matrix in a greater unitary matrix by the unitary dilation theorem \cite{halmos1951normal,mezher2023solving}.
\begin{align}
    U_{D} = \begin{pmatrix}
    D_{s} & \sqrt{I_{d_{1} \times d_{1}} - D_{s}(D_{s})^{\dagger}}\\
    \sqrt{I_{d_{2} \times d_{2}} - (D_{s})^{\dagger}D_{s}} & -(D_{s})^{\dagger}
    \end{pmatrix}\label{eq:unitary_dilation}
\end{align}
$U_{D} \in \mathbb{C}^{(d_{1}+d_{2}) \times (d_{1}+d_{2})}$ is a unitary matrix that can now be converted into a linear interferometer using efficient methods\footnote{This is unlike the challenge of decomposing unitary matrices into quantum gates for the gate-model.} \cite{reck1994experimental,clements2016optimal}. Since we know the rows and columns where $D_{s}$ has been embedded, we can directly focus on those rows and columns for this computation (namely, the first $d_2$ modes).
After the computation, we can do a procedure called post selection, where we filter out unwanted samples based on some criteria $E$, to then calculate the conditional probability of $\psi'$: $P(\psi'|\psi_{0},E)$. For our case, $E$ can be denoted by
\begin{align}
     E &= E_1 \wedge E_2 \label{eq:E}\\
    \text{where }E_{1} &= \bigwedge_{i=1}^{d_{1}} (n_{i}' \leq \tau) \label{eq:condE2}\\
    \text{and } E_{2} &= \bigwedge_{i=d_{1}+1}^{d_{1} + d_{2}} (n_{i}' = 0) \label{eq:condE1}
\end{align}
In other words, at the end of boson sampling, each of the first $d_{1}$ modes can have upto $\tau$ photons\footnote{Traditionally, $\tau=1$ in some versions of boson sampling.} but the next $d_2$ modes must have exactly 0 photons in them. This is because the first $d_{1}$ modes would represent the rows of $D_{s}$.

Let $\mathcal{M} = \{1,2,...,(d_{1}+d_{2})-1,d_{1}+d_{2}\}$ be a set of index positions from 1 to $d_{1}+d_{2}$. $\mathcal{C}' \subset M, |C'|=b$ is a set of the column index positions of our choice (from the first $d_2$ index positions). Essentially, $\mathcal{C}'$ is the set of columns of a submatrix $\beta'$ that serves as a candidate for a potential bicluster.

Thus, our starting state can be described as $\ket{\psi_0} = \ket{\mathcal{N}}$ where 
\begin{align}
    \mathcal{N} = \{n_i| (n_i = 1 \wedge i \in  \mathcal{C}') \vee (n_i = 0 \wedge i \in  \mathcal{M}-\mathcal{C}') \}\label{eq:N}
\end{align}
In other words, we generate a photon in the modes specified by $\mathcal{C}'$. After boson sampling, we extract the set of selected rows $\mathcal{R}'$ from the samples of the state with the highest probability $P(\psi'|\psi_{0},E)$.

Once we have $\mathcal{R}'$ and $\mathcal{C}'$, we can construct our candidate bicluster $\beta'$. We can then evaluate the quality of this candidate by a cost function $f()$ like the permanent \footnote{Since calculating the permanent is \#P-Hard, the use of boson sampling \cite{mezher2023solving} is suggested for this task as well.} or the matrix norm of $\beta'$. Figure \ref{fig:boson_sampling_flowchart} represents a visual version of this workflow.

Repeating this process for different choices of $\mathcal{C}'$ may give us different choices of $\mathcal{R}'$. And at the end, the candidate $\beta'$ with the best cost function score (largest permanent or norm) would be considered as a bicluster $\beta$ with columns $\mathcal{C}$ and $\mathcal{R}$. The process of figuring out the best choice of $\mathcal{C}$ can be viewed as a blackbox optimization process. In our work, we use SA for this task.

Once a bicluster $\beta$ is chosen and added to the set $\mathcal{B}$, we make the values corresponding to its rows and columns equal to 0 in $D_{s}$ and repeat the entire process again. We can do this for $k$ times (if the goal is to get $k$ biclusters) or use some other criteria for stoppage. This heuristic is encapsulated in Algorithm \ref{alg:bs_bicluster} in detail.

\begin{algorithm}
\caption{Boson sampling for finding biclusters of size $b \times b$ (simulated annealing for col. selection)}\label{alg:bs_bicluster}
\begin{algorithmic}[1]
\Procedure{MAIN}{$D,num\_samples,b,k,\varepsilon,f(),E,T$} \Comment{$f()$:cost function, $k$:no. of biclusters, $\varepsilon$: termination criteria (satisfiability expression), $E:$ postselection criteria, $T$: annealing temperature schedule}
\State Create $D_{s} \leftarrow \frac{1}{s}D$ where $s = \sigma_{max}(D)$
\State Initialize $i \leftarrow 0$, set $\mathcal{B} \leftarrow \emptyset$ and a set $\mathcal{I} \leftarrow \emptyset$
\While {$i< k$ \textbf{and} $\varepsilon \neq \text{True}$}
    \State Get $\ \beta,\mathcal{C},\mathcal{R} \leftarrow$ FINDBICLUSTER\_SA($D_{s},b,num\_samples,f(),E,T$)
    \State Add bicluster $\beta$ to the set of all biclusters, $\mathcal{B} \leftarrow \mathcal{B} \cup \beta$
    \State Append $\mathcal{I} \leftarrow \mathcal{I} \cup \mathcal{R} \times \mathcal{C}$ \Comment{Here, $\times$: cartesian product}
    \State Set all elements in $D_{s}$ to 0 for positions described in $\mathcal{R}\times\mathcal{C}$
    \State $i \leftarrow i + 1$
\EndWhile
\State \textbf{return} $\mathcal{B},\mathcal{I}$

\EndProcedure
\Procedure{FINDBICLUSTER\_SA}
{$D_{s},b,num\_samples,f(),E,T$}
\State Randomly select $b$ columns and create a set $\mathcal{C'}$. Intialize $cost \leftarrow 0$
\Repeat
    \State Get next $t$ from annealing schedule $T$\label{line:SA_1}
    \State Get  $\mathcal{R}' \leftarrow $ GETROWS($D_{s},b,num\_samples,\mathcal{C'},E$)
    \State Create candidate bicluster $\beta'$ by taking elements from $D_{s}$ indexed by $\mathcal{R'} \times \mathcal{C'}$
    \State Evaluate $cost' \leftarrow f(\beta')$
    \State $\Delta cost \leftarrow cost' - cost$
   \If{$\Delta cost > 0 $ \textbf{or} $U(0,1)< exp(\frac{\Delta cost}{t})$ }\label{line:SA_2}
        \State $\beta \leftarrow \beta'$, $cost \leftarrow cost'$, $\mathcal{C} \leftarrow \mathcal{C}'$ and $\mathcal{R} \leftarrow \mathcal{R'}$
    \EndIf
    \State Generate a new $\mathcal{C}'$ based on the neighborhood of $\mathcal{C}$ \label{line:SA_3}
\Until{all $t$s in $T$ are covered}
\State \textbf{return } $\beta,\mathcal{C},\mathcal{R}$
\EndProcedure
\Procedure{GETROWS}
{$D_{s},b,num\_samples,\mathcal{C}',E$}
\State Create unitary matrix $U_{D}$ using $D_{s}$ by Eqn(\ref{eq:unitary_dilation})
\State Prepare state $\ket{\psi_{0}} = \ket{\mathcal{N}}$ by using $\mathcal{C}'$ and Eqn(\ref{eq:N})
\State Do boson sampling for $num\_samples$ times
\State Get the state $\ket{\psi'}$ with the highest observed $P(\psi'|\psi_{0},E)$ where E is defined by Eqn(\ref{eq:E})
\State Extract the set of rows $\mathcal{R}'$ from $\ket{\psi'}$
\State \textbf{return } $\mathcal{R}'$
\EndProcedure
\end{algorithmic}
\end{algorithm}

\subsection{Dealing with rectangular biclusters}

While this work is primarily concerned with square biclusters, rectangular biclusters can also be preprocessed to be accommodated in this scheme. This step would essentially involve padding the matrix with columns of 1s.

We begin here with the precursor to matrix $D$; the matrix $D' \in \mathbb{R}^{d'_{1} \times d'_{2}}$ with each value in the interval $[0,1]$. We will describe a scheme that would allow us to find a bicluster $\beta \in \mathbb{R}^{b_1 \times b_2}$ with $b_1 > b_2$. This preprocessing method is restricted to finding biclusters where number of columns are fewer than the number of rows \footnote{$D'$ can have dimensions where $d_1=d_2$ or $d2 > d1$ or $d1 > d2$.}. So if our desired rectangular bicluster has $b_2 > b_1$, then we would need to start with $(D')^T$ such that we would be searching for $\beta^T \in \mathbb{R}^{b_2 \times b_1}$.

The general approach is to pad matrix $D'$ with $\Delta b$ columns\footnote{ $\Delta b = |b_1 - b_2|$.} of all 1s to create matrix $D$. 
This is done to have `anchored' columns $\mathcal{C}_{anchor}, |\mathcal{C}_{anchor}|= \Delta b$ where we send 1 photon each in their corresponding modes. The rest of the columns $\mathcal{C}_{choice}, |\mathcal{C}_{choice}|= b_2$ would be the actual choices for the columns of our bicluster (picked from the first $d_2$ modes). Together they would make $\mathcal{C}'$ (where $\mathcal{C}' \leftarrow \mathcal{C}_{choice} \cup \mathcal{C}_{anchor}$)

By sending photons through modes of `anchored' columns, we can make them available for row selection (after the computation) while still keeping track of a smaller number of columns for our actual bicluster. Since the anchored columns all have 1s, they would have a constant effect on the value of the permanent (even after scaling down by $s$) and can be ignored in the final creation of the bicluster. Algorithm \ref{alg:bs_preprocessing} describes this preprocessing process in detail.
\begin{algorithm}
\caption{Preprocessing data for finding rectangular biclusters (boson sampling)}\label{alg:bs_preprocessing}
\begin{algorithmic}[1]
\Procedure{MAIN}{$D',d'_1,d'_2,b_1,b_2$} 
\State Initialize variable $transpose\_flag = \text{False}$
\If{$b_2 > b_1$}
    \State Take its transpose $D' \leftarrow (D')^{T}$
    \State Swap values in $b_1$ and $b_2$
    \State Set $transpose\_flag \leftarrow \text{True}$ \Comment{This lets users know if they should transpose their biclusters}
\EndIf
\State Compute $\Delta b \leftarrow b_1 - b_2$
\State Create matrix $A_{1}$ of all 1s of size $d_1 \times \Delta b$, Create $D \leftarrow \begin{pmatrix} D' & A_{1} \end{pmatrix}$, where $D \in \mathbb{R}^{d_1 \times d_2}$ with $d_{1} \leftarrow d'_{1}$ and $ d_{2} \leftarrow d'_{2} +\Delta b$

\State \textbf{return } $D,d_{1},d_{2}, transpose\_flag$ 
\EndProcedure
\end{algorithmic}
\end{algorithm}

\begin{figure}
  \centering
   \includegraphics[scale=0.53]{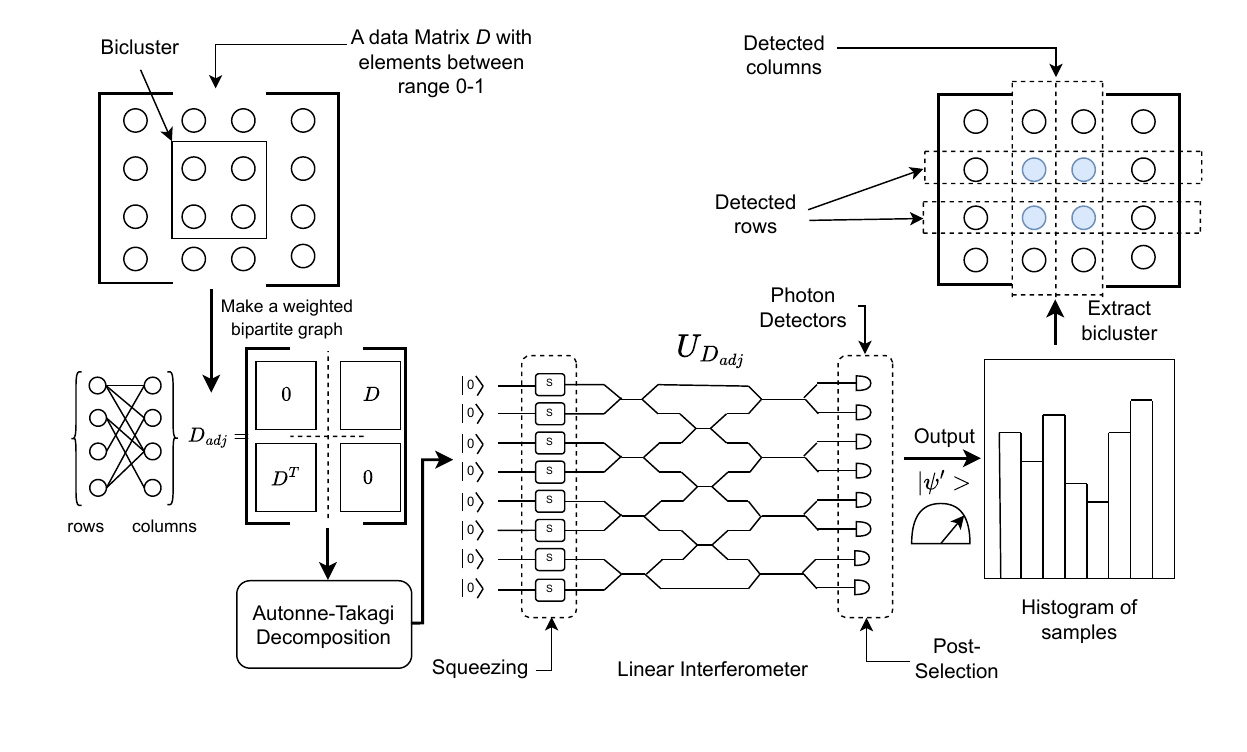}  
\caption{Workflow of the Gaussian boson sampling (GBS) approach for biclustering. Here, GBS returns the rows as well as the columns of potential biclusters (ideally with large hafnian or torontonian values). This figure is for illustrative purposes only.}
\label{fig:GBS_flowchart}
\end{figure}
\section{Gaussian boson sampling (GBS) for biclustering}
The process of using GBS for the task of biclustering is fairly straightforward. Additional preprocessing and post processing can be added to this method to make it more efficient, but we are going to focus on the basics of this application in our work.

Here, we will first convert our dataset $D \in \mathbb{R}^{d_1 \times d_2}$ (assumed to have values in the range $[0,1]$) into an adjacency matrix $D_{adj} \in \mathbb{R}^{(d_1 + d_2) \times (d_1+d_2)}$ by the following transformation
\begin{align}
    D_{adj} = \begin{pmatrix} 0 & D\\ D^T & 0 \end{pmatrix} 
\end{align}

This is done to prepare the matrix for the Autonne-Takagi decomposition\footnote{This process only works on symmetric matrices and $D_{adj}$ is one.} from which we will get the unitary matrix $U_{D_{adj}}$
\begin{align}
    D_{adj} = U_{D_{adj}} \text{diag} (\lambda_1,\lambda_2,...,\lambda_{d_1 + d_2}) U_{D_{adj}}^T \label{eq:autonne}
\end{align}

The $\lambda_{i}$s are used in the calculation of the squeezed states\footnote{Here, $\lambda$s are not to be taken as the eigenvalues of $D_{adj}$.}. Here it is assumed that the matrix $D_{adj}$ is scaled with the parameter $c > 0$ to make sure that $0 \leq \lambda_{i}\leq 1$. Together $c$ and $\lambda_{i}s$ are used in calculating the squeezing parameters $r_{i} = \tanh^{-1} (c\lambda_{i})$. They are also related to the mean number of photons to be generated by the following equation
\begin{align}
\overline{n} = \sum_{i=1}^{d_{1}+d_{2}}\frac{(c\lambda_{i})^{2}}{1- (c\lambda_{i})^{2}}
\end{align}

The unitary matrix $U_{D_{adj}}$ would then be converted to a linear interferometer and the squeezing parameters $r_i$ would be applied to their respective modes. After the computation, the columns $\mathcal{C}'$ and rows $\mathcal{R}'$ can be extracted from their corresponding modes in $\ket{\psi'}$ where one or more photons were detected.

Each GBS sample gives back a candidate bicluster $\beta'$. These candidate biclusters can be evaluated by a function $f()$ such as a matrix norm, based on which (and a threshold value) it can be accepted or rejected. The values in $D$ corresponding to the accepted bicluster are then set to zero\footnote{In this work, we only deal with non-overlapping biclusters. For getting overlapping clusters, we would need a more refined approach that includes (i) not setting the values to 0 after extracting a bicluster and (ii) a larger number of samples.}. The process can then repeat $k$ times or until a termination criteria $\varepsilon$ is reached.

Finding biclusters can be represented as the process of locating dense sub-graphs in a bipartite graph \cite{karim2019bicluso}. Searching for $N$-dense subgraphs using GBS is a topic that is being actively explored \cite{arrazola2018using,solomons2023effect}. The major restriction is that the sub-graphs (or biclusters) that GBS finds will have a even numbered dimension ($N$ is even). In order to find biclusters for an odd numbered $N$, and for biclusters where specific dimensions $b_1 \times b_2$ are needed, you would need to post-process the results\footnote{Since in GBS, both the rows and columns are observed after the computation.}. Figure \ref{fig:GBS_flowchart} illustrates a general idea of how GBS can be used to find biclusters and algorithm \ref{alg:gbs_bicluster} details this heuristic more formally.

\begin{algorithm}[!h]
\caption{Gaussian boson sampling for finding biclusters (non-overlapping)}\label{alg:gbs_bicluster}
\begin{algorithmic}[1]
\Procedure{MAIN}{$D,num\_samples,\varepsilon,\overline{n},k,f()$}
\State Initialize $\mathcal{B} \leftarrow \emptyset$, $\mathcal{I} \leftarrow \emptyset$ and $i \leftarrow 0$
\While {$i < k$ and $\varepsilon \neq \text{True}$ }
\State Create $D_{adj} \leftarrow \begin{pmatrix} 0 & D\\ D^T & 0 \end{pmatrix} $
\State Figure out scaling parameter $c$ based on $D_{adj}$ and $\overline{n}$
\State Scale the matrix $D_{adj} \leftarrow c D_{adj}$
\State Do Autonne-Takagi decomposition to get $U_{D_{adj}}$ and $\Lambda = \{\lambda_1,\lambda_2,...,\lambda_{(d_1+d_2)}\}$
\State Get a set of squeezing parameters $R = \{r_1,r_2,...,r_{(d_1+d_2)}\}$ by $r_{i} \leftarrow \tanh^{-1} (c\lambda_{i}), \forall r_{i} \in R$

    \State Do GBS using $U_{D_{adj}}$, $R$ to get $num\_samples$ number of samples
    \While{ samples remain to be processed}
        \State Get the candidate rows $\mathcal{R}'$ and columns $\mathcal{C}'$ from the current sample.
        \State Get the corresponding bicluster candidate $\beta'$ and 
        \If{ $\beta'$ has a good value of $f()$}
            \State $\beta \leftarrow \beta'$, $\mathcal{R} \leftarrow \mathcal{R}'$ and $\mathcal{C} \leftarrow \mathcal{C}'$ 
            \State Store it in $\mathcal{B} \leftarrow \mathcal{B} \cup \beta$ and append $\mathcal{I} \leftarrow \mathcal{R} \times \mathcal{C}$
            \State Set all elements in $D$ to 0 for positions described in $\mathcal{R}\times\mathcal{C}$
            \State $ i \leftarrow i + 1$
        \EndIf
    \EndWhile
\EndWhile

\State \textbf{return } $\mathcal{B}, \mathcal{I}$
\EndProcedure
\end{algorithmic}
\end{algorithm}

\section{Simulations and Results}\label{sec:sim_n_res}
In order to do preliminary validation of our ideas, we performed a few simulations to get an initial estimate of our proposed approaches. All our simulations were done on classical computers with most of them being done on a high performance server with a {\tt AMD Ryzen Threadripper 3970X 32-Core} processor with a memory size of {\tt $\sim$256.7 GB}. The programs were written in the Python programming language and the simulation packages used were {\tt Perceval} for boson sampling and {\tt Strawberry Fields} for GBS. All our simulations were noiseless.

We also wanted to experiment on real photonic quantum processing units (QPU) but at the time of the project, there were no publicly available QPUs that could handle problems of a large enough size. With the preliminary simulations, our aim was to select problems that are not intractable to simulate\footnote{While not being impossible, the simulation process was still slow for the resources we had.}, but are also non-trivial in nature.

We took a total of four problems : two for boson sampling and two for GBS with $12 \times 12$ sized datasets. The reason for choosing the above dimensions was to have a matrix in which biclusters of non-trivial sizes \footnote{$4 \times 4$ and $6 \times 6$.} could be embedded in. In other words, like mentioned above, because this is the first work of its kind, the chosen problems were designed to have a balance of simplicity and non-triviality. Another thing to note is that our simulations were based on a simplified version of the proposed heuristics in algorithms \ref{alg:bs_bicluster} and \ref{alg:gbs_bicluster}. This was done (i) partially due to the nature of these problems being simple and straightforward but also because of (ii) the simulation times involved. The exact details are mentioned in the following sections. We hope that the insights from these simulations would be useful for future implementation on real photonic hardware.
\subsection{Boson sampling-problem 1}\label{sec:bs_exp1}
\subsubsection{Setup}
In our first simulation, we are going to have the following assumptions:
\begin{enumerate}
    \item There is only one bicluster $\beta$ in our dataset $D$.
    \item We know the set of columns $\mathcal{C}$ for the $\beta$.
\end{enumerate}
Our task here is to use boson sampling with input created from $\mathcal{C}$ to calculate the probability of observing the correct set of rows $\mathcal{R}$.
For this, we created a $6\times6$ matrix that will act as our bicluster $\beta$ that has values randomly sampled from $\{0.7,0.8,0.9\}$. Then we created $D^{(1)},D^{(2)},...,D^{(5)}$ datasets of size $12\times12$ where we embed $\beta$ at $\mathcal{C}, \mathcal{R} = \{4,5,6,7,8,9\}$. All other elements in the datasets have values from $0$ to $0.1 \times \alpha$:
\begin{align}
 D^{(\alpha)}_{\begin{subarray}{l} ij \\
    i\notin \mathcal{R}, j\notin \mathcal{C}\end{subarray}} \in \{(0.1)h | 0 \leq h \leq \alpha,h \in\mathbb{Z}\}
\end{align}
We created one more matrix $D^{(6)}$ in which we embed a $6 \times 6$ bicluster of all 1s (for the same $\mathcal{R}$ and $\mathcal{C}$). The rest of the values are all 0s. This is done to study boson sampling's performance for binary matrices, albeit a very simple one in this case.

For each dataset and our fixed $\mathcal{C}$ we performed boson sampling with $10^5$ and $10^6$ samples. We then analyzed the success probability $\mathcal{P}$ of getting our ideal bicluster for 3 different conditions:
\begin{enumerate}
    \item No postselection (raw results)
    \item Postselection as defined by Eqn(\ref{eq:E}) with $\tau=1$.
    \item Postselection as defined by Eqn(\ref{eq:E}) with $\tau=3$.
\end{enumerate}
For points 1 and 2 in the list above, in order to estimate $\mathcal{P}$, a success is to observe if the modes corresponding to $\mathcal{R}$ receive 1 photon each. Here, the success probability $\mathcal{P}$ is based on finding (single) photons in all rows described in $\mathcal{R}$ and no others. For \#3, we are fine with up to three photons in the same mode, but we only count each unique row once. Since the number of photons are limited, in this case, success is defined as finding photons in a subset of the rows $\mathcal{R}$ of the ideal bicluster, but no others.
\subsubsection{Results and discussion}

\begin{table}
\centering
\begin{tabular}{ |c|c|c|c|c|c|c|c|c|c|c| } 
 \hline
  \multicolumn{1}{|c|}{\textbf{Dataset}} & \multicolumn{1}{c|}{\textbf{Heatmap}} & \multicolumn{3}{c|}{\textbf{no-postselect}}& \multicolumn{3}{c|}{\textbf{postselect $\tau=1$}} & \multicolumn{3}{c|}{\textbf{postselect $\tau= 3$}} \\
 \hline
 & & \textbf{Num.} & \textbf{Den.} & \textbf{$\mathcal{P}$} & \textbf{Num.} & \textbf{Den.} & \textbf{$\mathcal{P}$} & \textbf{Num.} & \textbf{Den.} & \textbf{$\mathcal{P}$} \\
\cline{3-11}
$D^{(1)}$ & \includegraphics[scale=0.17]{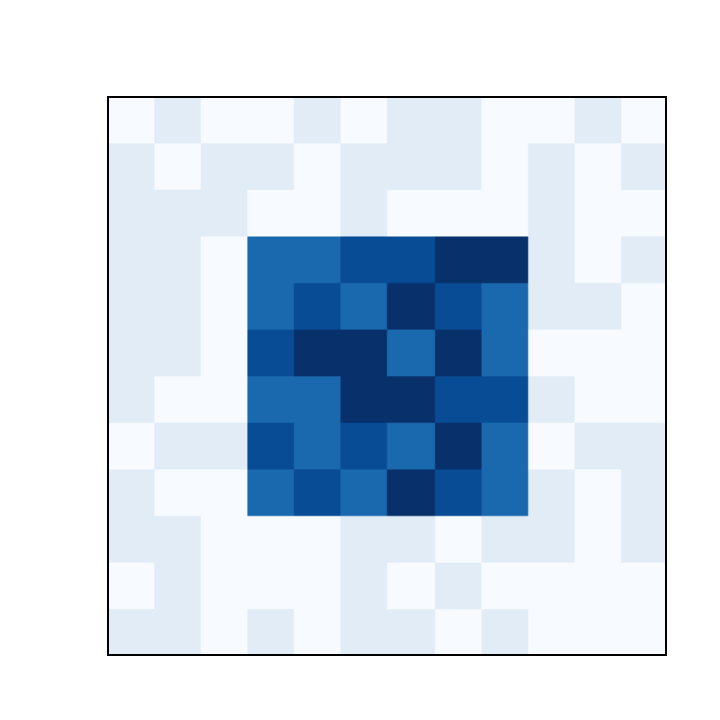} & $219$ & $10^6$ & $2.1\times10^{-4}$ & $219$ & $245$ & $0.893$ & $13269$ & $13593$ & $0.976$ \\
\hline
 
$D^{(2)}$ & \includegraphics[scale=0.17]{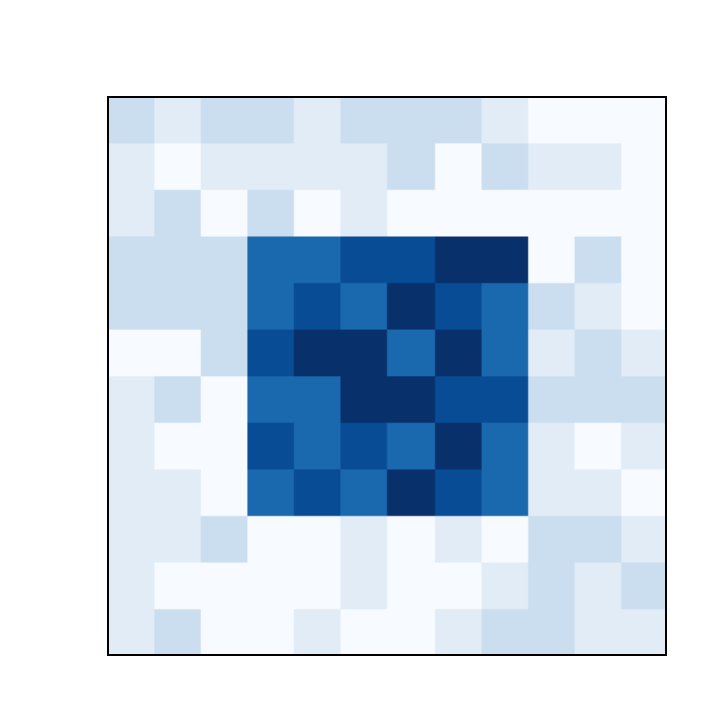} & $185$ & $10^{6}$ & $1.8\times10^{-4}$ & $185$ & $262$ & $0.706$ & $11323$ & $12289$ & $0.921$ \\
\hline
$D^{(3)}$ & \includegraphics[scale=0.17]{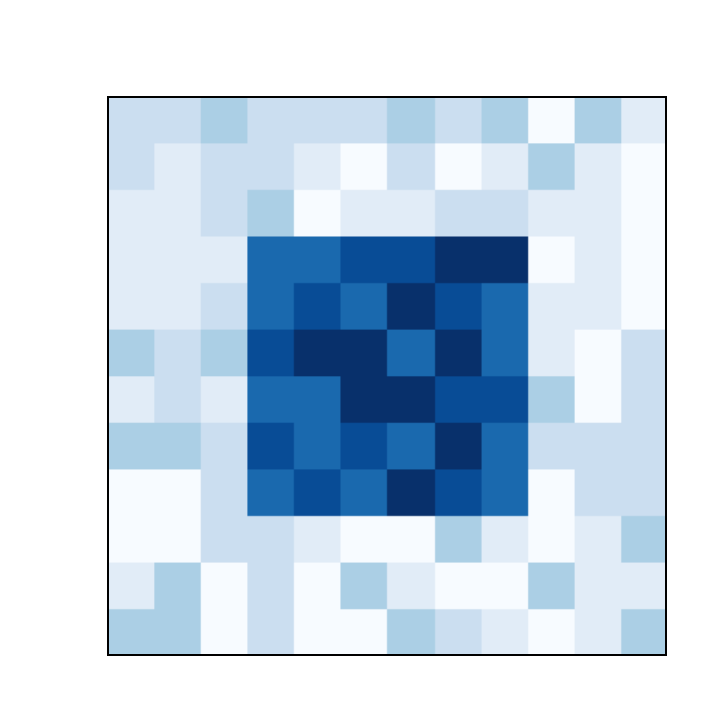} & $152$ & $10^{6}$ & $1.5\times10^{-4}$ & $152$ & $366$ & $0.415$ & $9101$ & $11277$ & $0.807$ \\
\hline
$D^{(4)}$ & \includegraphics[scale=0.17]{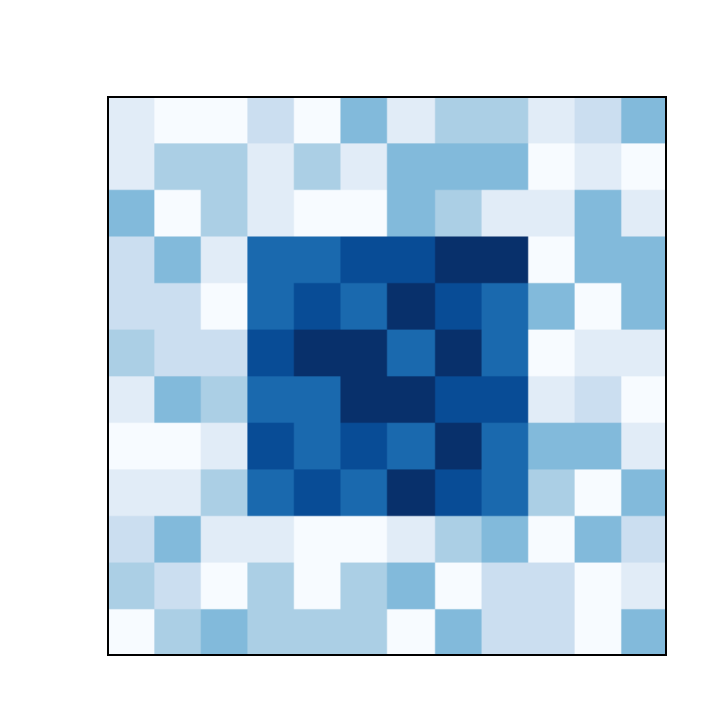} & $101$ & $10^6$ & $1\times10^{-4}$ & $101$ & $514$ & $0.196$ & $5909$ & $9130$ & $0.647$ \\
\hline
 $D^{(5)}$ & \includegraphics[scale=0.17]{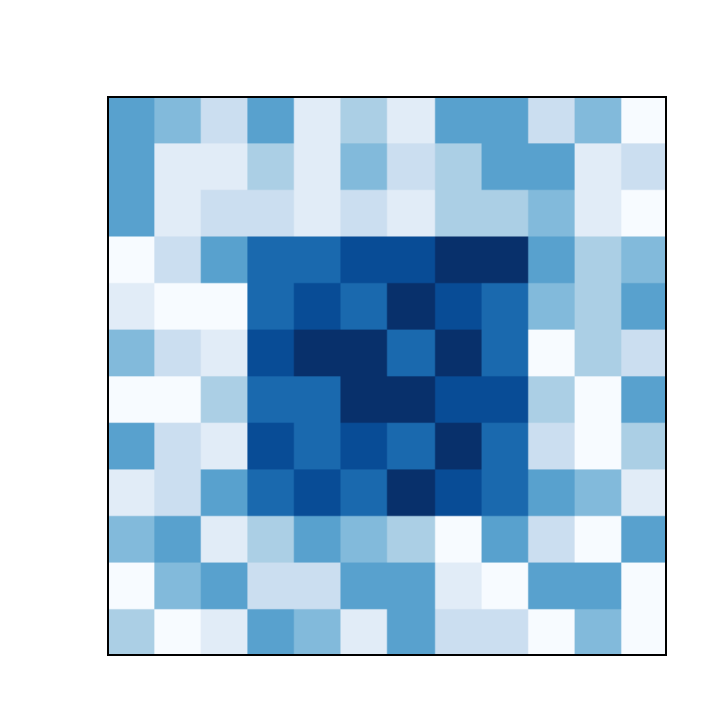} & $49$ & $10^{6}$ & $4.9\times10^{-5}$ & $49$ & $559$ & $0.087$ & $3007$ & $6618$ & $0.454$ \\
 \hline
 $D^{(6)}$ & \includegraphics[scale=0.17]{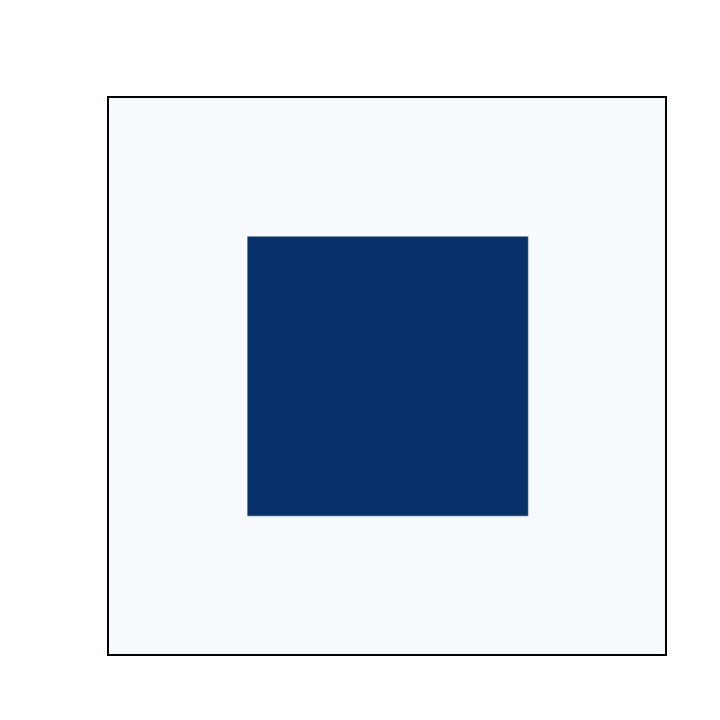} & $255$ & $10^{6}$ & $2.5\times10^{-4}$ & $255$ & $255$ & $1.0$ & $14686$ & $14686$ & $1.0$ \\
\hline
\end{tabular}
\caption{Success probabilities $\mathcal{P}$ for boson sampling-problem 1 of datasets $D^{(1)}$ to $D^{(6)}$, when calculated for 1 million samples. The heatmaps depicts the areas of the table where the values are the highest (dark blue) and lowest (white or very light blue). For each of the three settings (no-postselection, postselection with $\tau=1$ and postselection with $\tau=3$), the numerator (no. of samples containing the ideal result), denominator (no. of samples filtered through the postselection criteria $E$) and $\mathcal{P}$ (numerator/denominator) is listed. }\label{table:bs_exp1}
\end{table}

The results of the simulations on our first problem can be best described by Table \ref{table:bs_exp1}. This table shows us the results for cases where 1 million samples were taken. It should be noted that our results for 100k samples were comparable, but we chose to publish the former since the calculated probabilities would be closer to their exact values (due to the larger sample size).

The first observation we can make is the effect that the values in $(D^{(\alpha)}_{ij})_{i\notin \mathcal{R}, j\notin \mathcal{C}}$ can have on finding the bicluster $\beta$. As the values in the other parts of the dataset go up (even if its lower than the values in the rows of $\beta$), the probability of those rows to be sampled also goes up (even if the probability of finding the rows of $\beta$ may still be the highest). For example, for the data of postselect $\tau=1$, we can see how the results degrade when going from $0.89$ for $D^{(1)}$ to $0.087$ for $D^{(5)}$. This seems to suggest that such a technique would do better if there is enough difference in the magnitude of the values of the bicluster and the rest of the dataset. This is similar to how in quantum annealing, finding the ground state solutions becomes worse when there are plenty of other good solutions in the landscape (in other words, when a landscape is not `rugged') \cite{king2019quantum}. Similarly, the equivalent of a lot of ruggedness in boson sampling terms would seem to be a larger difference in the magnitude of $\beta$ and $(D^{(\alpha)}_{ij})_{i\notin \mathcal{R}, j\notin \mathcal{C}}$.

The other important observation to make is the role postselection plays for this task. In the million samples, the majority of them have photons that may be (i) incident on modes $>d_1$ and/or (ii) aggregate in modes in quantities $>1$. After applying our strictest postselection criteria (postselect $\tau=1$ column), the samples that remain are only in the hundreds. The situation is somewhat better for our other postselection criteria (postselec $\tau=3$) where the samples that remain are in the thousands. Regardless, based on these simulations, it would appear that in order to even conduct meaningful postselection, we need a very large number of samples. Depending on the hardware equipment being used, the time taken to get one sample in a linear interferometer for boson sampling and GBS can be very fast \cite{madsen2022quantum,deng2023gaussian}.

Like mentioned before, the success probability for when postselection is done with $\tau=3$ is altered to be the probability of samples where one or more photons are received on a mode in $\mathcal{R}$\footnote{And also like mentioned before, if we were to extract rows from such a sample (to construct $\beta'$), we would only take the unique ones.}. Therefore, not only are we working with a higher denominator when $\tau=3$, but also a comparitively high numerator value. And while $\mathcal{P}$ does decrease from $D^{(1)}$ to $D^{(5)}$, it is higher than its $\tau=1$ counterpart. Essentially it indicates that, the traditional way to run boson sampling (with single photons in the input mode and a requirement of only allowing one photon in a mode at the output) may not be best suited for practical applications. Indeed, along with the lower probability, this stringent approach will face issues with noise once we would try and implement it on an actual photonic device \cite{brod2019photonic} (photon loss being a major concern among them).

Finally, the results for $D^{(6)}$ shows us how boson sampling may behave for a bicluster of all 1s with all other parts of the dataset containing zeros. Here we can see that the postselected $\mathcal{P}$ is observed to be 1. The success probabilities of a realistic binary dataset however, may not be expected to be so high as there would most likely be presence of 1s outside of the actual biclusters. While typical binary datasets would probably not have such a simple spread of values across the entire matrix, the value of our simple simulation is to show how extreme contrast in values can affect the probabilities. 

\subsection{Boson sampling-problem 2}\label{sec:bs_exp2}
\subsubsection{Setup}\label{sec:bs_exp2_setup}
For our second simulation, we use dataset $D^{(2)}$ from the first problem\footnote{$D^{(2)}$ was chosen over the others for a decent contrast of values in the dataset (useful for a good $\mathcal{P}$) while still being challenging enough for boson sampling.} but shuffle its rows and columns randomly to make it into a slightly more challenging problem. The objective is to find a $\beta \in \mathbb{R}^{6 \times 6}$ where we do not know its $\mathcal{R}$ or $\mathcal{C}$. Here, our assumption is that there is one and only one $6 \times 6$ bicluster in the dataset.

 In order to do this, we use algorithm \ref{alg:bs_bicluster} with the following parameters:
\begin{enumerate}
    \item $D = P_{R}(D^{(2)}P_{C})$ where $P_{R}$ and $P_{C}$ are the randomly generated row and column permutation matrices respectively.
    \item $num\_samples = 10^5, b =6$ and $k=1$
    \item $\varepsilon = \emptyset$, essentially we don't have any separate termination critera.
    \item $f()$ is the permanent
    \item $T$ is an exponential decay annealing schedule: by $T = \{t_{i} \vert t_{i} = t_{0}\big(\frac{t_{f}}{t_{0}}\big)^{i/p}, 0 \leq i \leq b-1, i \in \mathbb{Z} \}$ with $t_{0}=100$ and $t_{f} = 0.01$
\end{enumerate}
In each iteration of the SA process, the rows are chosen from the sample that appears the most after postselection. We use the postselection criteria $E$ as mentioned in Eqn(\ref{eq:E}) with initial $\tau=1$. If we do not find any samples that satisfy this criteria, we iteratively increase $\tau$ upto $b-1$. If no samples are selected even after this, we assign $\mathcal{R}'\leftarrow \emptyset$ and $cost'= -1000$. This means that the selection of $\mathcal{C}'$ was very bad and should not be accepted.
\begin{figure}
  \centering
   \includegraphics[scale=0.6]{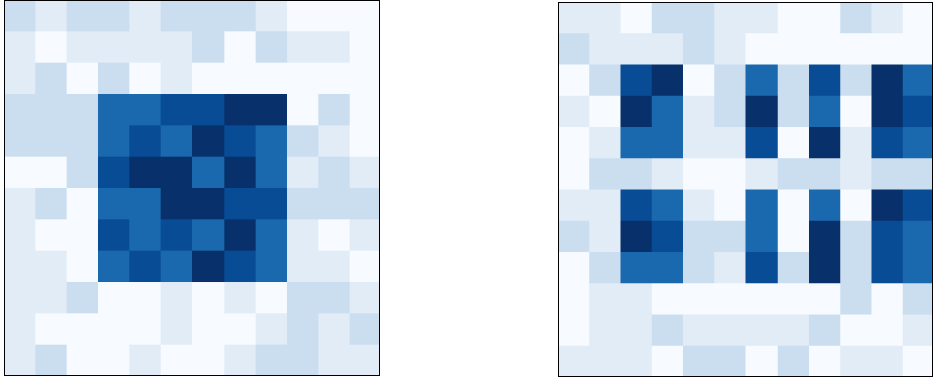}  
\caption{(RIGHT) Heatmap of the dataset used in boson sampling-problem 2. This problem was generated by taking $D^{(2)}$ from boson sampling-problem 1 whose heatmap is on the (LEFT) and then performing a random permutation on its rows and columns.}
\label{fig:bs_problem2}
\end{figure}
For samples where $\tau>1$, it means that $|\mathcal{R}'| \leq 6$. For these simulations, whenever $|\mathcal{R}'| < 6$, we chose to reduce the $|\mathcal{C}'|$ to match $|\mathcal{R}'|$ by dropping the columns in $\mathcal{C}'$ with the lowest individual L2 norms \footnote{Higher elements in a column would correlate with higher (vector) norm values.}, which will give us $\beta'$, our bicluster candidate. Essentially, for these simulations on problem 2, our bicluster can potentially be smaller than $6 \times 6$ ( but not rectangular) if a bicluster of the original size cannot be found. 

For SA, the neighborhood of $\mathcal{C}$ is defined as any column's index position in $D$ outside of $\mathcal{C}$ which can be swapped for a index position inside $\mathcal{C}$. In other words, the process of generating new candidate $\mathcal{C}'$ is as follows:
\begin{enumerate}
    \item Copy $\mathcal{C}' \leftarrow \mathcal{C}$
    \item Randomly select a column index position $i$ of $D$ where $i \in \mathcal{C}'$
    \item Randomly select a column index position $j$ of $D$ where $j \notin \mathcal{C}'$
    \item Delete $i$ from $\mathcal{C}'$ : $\mathcal{C}' \leftarrow \mathcal{C}' - \{i\} $
    \item Add $j$ to $\mathcal{C}'$: $\mathcal{C}' \leftarrow \mathcal{C}' \cup j $
\end{enumerate}

We ran our simulations for $p$ steps that span over the entire annealing schedule $T$ for $p\in \{20,50,100,150,200\}$ for a hundred trials per $p$ (500 total). Each SA step only involves one Monte Carlo sweep (i.e only one iteration per temperature value in T).

\subsubsection{Results and discussion}\label{sec:bs_exp2_results}

\begin{table}[h]
\centering
\begin{tabular}{ |c|c|c|c|c|c| } 
 \hline
 \textbf{Anneal Steps $p$} & $20$ & $50$ & $100$ & $150$ & $200$ \\ 
\hline
\textbf{Success Probability $\mathcal{P}$} & $0.17$ & $0.34$ & $0.85$ & $0.95$ & $0.99$ \\  
 \hline
\end{tabular}
\caption{Approximate success probabilities $\mathcal{P}$ for Problem 2 of boson sampling for corresponding number of steps that simulated annealing (SA) was run for (boson sampling being used as a subroutine).}\label{table:bs_exp2}
\end{table}

The results for the simulation of problem 2 as encapsulated by Table \ref{table:bs_exp2} do show a favorable scaling of finding a square bicluster for the number of SA steps $p$ involved. Of course, since we have just considered one problem over here, a more thorough study in the future is warranted.

One of the most glaring challenges is the number of samples that we had to do for this process to work. As mentioned previously, we ran boson sampling for $10^{5}$ times per each anneal step. This also constrained us in limiting the number of times we run the simulation per $p$. However, it should also be kept in mind that actual devices are going to be way faster than the simulation work we did. A different black box technique to SA could also be considered.

\subsection{Gaussian boson sampling-problem 1}\label{sec:gbs_exp1}
\subsubsection{Setup}\label{sec:gbs_exp1_setup}

Since the approach of finding biclusters using GBS yields us both the rows and columns of the bicluster simultaneously, for our first experiment, we consider the same dataset $D$ as the one for boson sampling-problem 2 (see figure \ref{fig:bs_problem2}). We also consider a binary version $D_{bin}$ that makes all values $\geq 0.7$ from the original equal to one and zero if otherwise. Our goal is to find the same $6 \times 6$ bicluster as the one in section \ref{sec:bs_exp2}.

In GBS, because we cannot control the exact number of photons we send through the circuit, the mean number of photons $\overline{n}$ per mode (or just $\overline{n}$ per mode) is an important hyperparameter for the GBS process. As we can see from Eqn (\ref{eq:probability_gbs}), having either too low or too high a value of $\overline{n}$ per mode may affect the results in a negative way\footnote{Too low : photons may not be generated and/or lost to photon loss. Too high: the probability of finding good biclusters will go down factorially.}. 

Our simulations were done with 1,2,4,6 and 10 mean photons per mode. We took $10^{4}$ samples for each setting of $\overline{n}$ per mode. The primary reason for taking fewer samples than boson sampling is that the GBS approach seems to require a smaller number of samples for other applications \cite{arrazola2018using,bonaldi2023boost}. The secondary reason is that the cost of simulating a GBS process is far larger for a 24-mode linear interfermetor than for a boson sampling circuit of the same size. This is because a classical process has to keep track of squeezed light states across a large number of modes in GBS.

After the process, photons in the first $d_{1}$ modes would represent the rows that have been selected for a candidate bicluster and the next $d_{2}$ modes would represent the selected columns for the same. For measurement, we used the threshold detection process as implemented in the {\tt strawberryfields} package. So even if we had more than one photon in a mode, we would still count that row/column only once. Finally, once we decipher the proposed biclusters from the samples, we then compare them against the correct result in order to calculate the success probability $\mathcal{P}$.

\subsubsection{Results and discussion}\label{sec:gbs_exp1_results}

\begin{table}[h]
\centering
\begin{tabular}{ |c|c|c|c|c|c| } 
 \hline
 \textbf{$\overline{n}$ per mode} & $1$ & $2$ & $4$ & $6$ & $10$ \\ 
\hline
\textbf{No. of correct samples} & $735$ & $1038$ & $1099$ & $945$ & $701$ \\ 
\hline
\textbf{Success Probability $\mathcal{P}$} & $0.0735$ & $0.1038$ & $0.1099$ & $0.0945$ & $0.0701$ \\  
 \hline
\end{tabular}
\caption{Approximate success probabilities $\mathcal{P}$ for problem 1 of Gaussian boson sampling (GBS) for corresponding mean number of photons ($\overline{n}$) per mode. This is for the real valued dataset $D$.}\label{table:gbs_exp1_real}
\end{table}

\begin{table}[h]
\centering
\begin{tabular}{ |c|c|c|c|c|c| } 
 \hline
 \textbf{$\overline{n}$ per mode} & $1$ & $2$ & $4$ & $6$ & $10$ \\ 
\hline
\textbf{No. of correct samples} & $1949$ & $4125$ & $6180$ & $7305$ & $8234$ \\ 
\hline
\textbf{Success Probability $\mathcal{P}$} & $0.1949$ & $0.4125$ & $0.6180$ & $0.7305$ & $0.8234$ \\  
 \hline
\end{tabular}
\caption{Approximate success probabilities $\mathcal{P}$ for problem 1 of Gaussian boson sampling (GBS) for corresponding mean number of photons ($\overline{n}$) per mode. This is for the binary dataset $D_{bin}$.}\label{table:gbs_exp1_bin}
\end{table}
From tables \ref{table:gbs_exp1_real} and \ref{table:gbs_exp1_bin}, the very first thing the data seems to suggest, is that GBS, like boson sampling, seems more effective at the task of finding binary biclusters. But we don't need success probabilities $\geq 2/3$ in order for GBS to be useful. As long as we have samples from which the part or whole of a bicluster can be extracted\footnote{This may be done by evaluating the candidate bicluster of each sample using a cost metric $f()$ like a matrix norm, and selecting the bicluster with the highest cost value.}, there may be some utility of GBS for this application. Another thing to note is that the success probability of a random sampling procedure to find the correct bicluster is in the order of\footnote{There are a total of 24 modes. For simplifying the calculation, consider the values the modes can take to be either 0 or 1 (photons are either there or not). Total number of possible solutions (for a $6 \times 6$ bicluster): ${26 \choose 6} = 134596$ and probability of sampling the best solution (assuming there is only one) becomes $1/134596\approx 7.43 \times 10^{-6}$. } $10^{-6}$ , which is lower than the success probabilities from GBS. It should also be mentioned that except in one case (real valued dataset, $\overline{n}$ per mode $=1$), the correct samples were in the majority (i.e. the statistical mode).

The other important observations with GBS are that (i) we figure the rows and columns of the bicluster simultaneously and (ii) we may be able to get meaningful results with fewer number of samples when compared to our approach with boson sampling. Of course, further studies and a thorough comparison with classical methods is needed to make stronger claims.

The final observation for this problem is how the different datasets performed for different values of $\overline{n}$ per mode. For $D$, 4 mean photons per mode gave us the best success probability and for $D_{bin}$ it was 10 (from the limited information we have). Regardless, the initial data indicates that the best value for $\overline{n}$ per mode could be different from problem to problem.

\subsection{Gaussian boson sampling-problem 2}\label{sec:gbs_exp2}
\subsubsection{Setup}\label{sec:gbs_exp2_setup}
\begin{figure}
  \centering
   \includegraphics[scale=0.6]{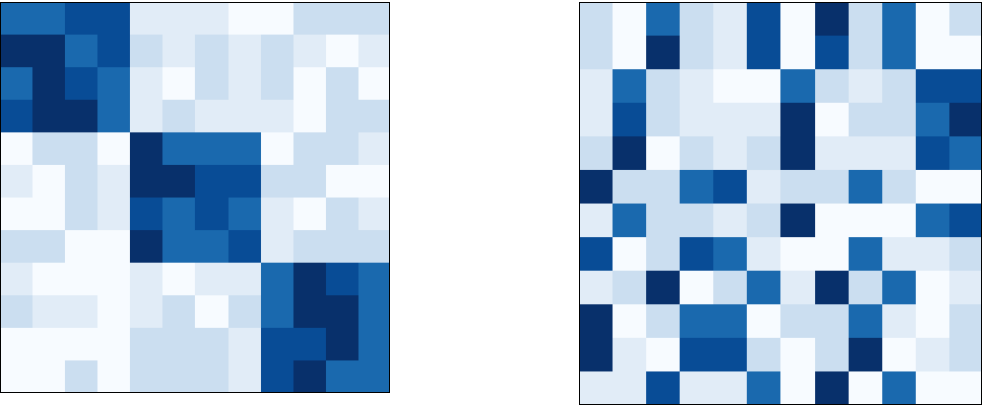}  
\caption{(RIGHT) Heatmap of the dataset with 3 biclusters used in GBS-problem 2. This problem was generated by first creating the dataset whose heatmap is on the (LEFT) and then performing a random permutation on its rows and columns. For further details, refer section \ref{sec:gbs_exp2_setup}.}
\label{fig:gbs_problem2}
\end{figure}
For the second simulation of GBS, we chose a problem that has three biclusters $\beta_{1},\beta_{2}, \beta_{3} \in \mathbb{R}^{4\times4}$ that has values in the interval $[0.7, 0.9]$. These three biclusters are placed along the diagonal of a $12 \times 12$ dataset $D$, the rest of which has values $\leq 0.2$.
After this, a random permutation was applied on the rows and columns of the dataset to make the problem more challenging and realistic\footnote{Since one may not expect to find an ideal bicluster located entirely contiguosly across the dataset.}. Figure \ref{fig:gbs_problem2}, shows the heatmaps of the dataset during its initial and final creation phases (LEFT and RIGHT parts of the figure respectively). Finally a binary dataset $D_{bin}$ was created by the same criterion as mentioned in section \ref{sec:gbs_exp1_setup}.

The aim of this simulation was to see how effective GBS would be for detecting multiple biclusters. We chose 2 as the $\overline{n}$ per mode value after doing some preliminary simulations of different values (for small number of samples). Here, we applied a simplified version of the heuristic from algorithm \ref{alg:gbs_bicluster} to reduce the number of simulations we need to do while still being able to extract useful results. Essentially, after performing GBS once for $10^{4}$ samples, we remove two of the correct biclusters that have the highest counts in all the samples by replacing their values with 0s in $D$ (or $D_{bin}$). The GBS process is then done once more for the same number of samples ($10^4$) to see the change in measuring the last remaining bicluster.
\subsubsection{Results and discussion}\label{sec:gbs_exp2_results}

After applying GBS to our $12 \times 12$ datasets, we notice that the number of times the correct biclusters were successfully observed were in single digits for the real-valued dataset and in the low double-digits for the binary one (see table \ref{table:gbs_exp2_1}). At first glance, these results look very underwhelming. But we would like to bring to attention two points: firstly, even for the real-valued dataset, these results are still orders of magnitude better than results we may expect from random sampling\footnote{As mentioned in section \ref{sec:gbs_exp1_results}, random sampling has a probability in the order of $10^{-6}$, the worst of our GBS results suggest probabilities in the order of$10^{-4}$.}. This is corroborated by the 2013 theoretical work\footnote{While that work is based on traditional boson sampling, it would also broadly apply for the comparison in question.} by Aaronson and Arkhipov \cite{aaronson2013bosonsampling}.

Secondly, the standard way that the GBS process is set up is as a fixed circuit that runs in constant time (for a given size), unlike a gate-model quantum algorithm or a quantum annealing process. Thus, to an extent, taking more samples for GBS is more tolerable than taking more samples in the other types of models \footnote{Of course, in order to make a stronger case for GBS to be used for this application, it would still have to yield meaningful results that have to be better than the alternatives in some way(s). What we are suggesting is that our results do not take GBS out of contention for being a candidate for biclustering.}.
\begin{table}[h]
\centering
\begin{tabular}{ |c|c|c|c|} 
 \hline
 \multirow{ 2}{*}{\textbf{Dataset}} & \multicolumn{3}{c|}{\textbf{No. of correct samples}} \\ 
\cline{2-4}
 & \textbf{Bicluster $\beta_{1}$} & \textbf{Bicluster $\beta_{2}$} & \textbf{Bicluster $\beta_{3}$} \\
 \hline
 $D$ & 3 & 5 & 6 \\
 \hline
 $D_{bin}$ & 41 & 34 & 24\\
 \hline
\end{tabular}
\caption{Results of GBS-problem 2 for $10^4$ samples for real-valued and binary datasets ($D$ and $D_{bin}$ respectively. We can see that GBS performs better for $D_{bin}$ than for $D$.}\label{table:gbs_exp2_1}
\end{table}

\begin{table}[h]
\centering
\begin{tabular}{ |c|c|c|c|} 
 \hline
 \multirow{ 2}{*}{\textbf{Dataset}} & \multicolumn{3}{c|}{\textbf{No. of correct samples}} \\ 
\cline{2-4}
 & \textbf{Bicluster $\beta_{1}$} & \textbf{Bicluster $\beta_{2}$} & \textbf{Bicluster $\beta_{3}$} \\
 \hline
 $D$ & 1030 & 0 & 0 \\
 \hline
 $D_{bin}$ & 0 & 0 & 5966\\
 \hline
\end{tabular}
\caption{Results on GBS-problem 2 datasets after the top two results in each category were removed (see table \ref{table:gbs_exp2_1}). For $D$, it was $\beta_2$ and $\beta_3$ and for $D_{bin}$ it was $\beta_{1}$ and $\beta_{2}$. }\label{table:gbs_exp2_2}
\end{table}

Table \ref{table:gbs_exp2_2} shows the results of GBS once the two most observed biclusters (from the previous run) were removed from the datasets. Here, we see a significant increase of observing the remaining bicluster, with the largest rise seen for the binary dataset. This suggests that a full-fledged GBS heuristic that detects and removes biclusters iteratively may improve its chances for finding one or more biclusters in the $i$\textsuperscript{th} iteration than in the $(i-1)$\textsuperscript{th} iteration. Though further studies are needed before we can say more, we hope that the insight from this result is useful to the research community at large.

\subsection{A short summary of the results}\label{sec:summary}
Following is the summary of the results from our simulations:
\begin{enumerate}
    \item Both boson sampling and GBS can be applied to real-valued and binary datasets to detect biclusters within larger datasets.
    \item Both boson sampling and GBS perform better when the contrast between the values that are inside and outside the bicluster is large.
    \item The GBS approach seems to produce comparable or better results for fewer number of samples (against our current way of performing boson sampling).
    \item The GBS approach provides us with both the rows and columns of the bicluster simultaneously.
    \item If multiple biclusters are present within a dataset, then their chances of being detected goes down. We hope this can be somewhat alleviated in the future by the use of preprocessing and postprocessing methods\footnote{E.g. for preprocessing, you can use a threshold value to generate a binary version of a real-valued dataset \cite{bonaldi2023boost}. Similarly, postprocessing methods that build on the raw solutions in order to produce better solutions can also be considered \cite{arrazola2018using}.} to extract meaningful results.
\end{enumerate}

\section{Future work}
Based on the outcomes of section \ref{sec:sim_n_res}, we can now comment on the potential future work that would help to better understand the utility of these restricted models of quantum computation (enabled by photonics) for the problems of biclustering.

\begin{enumerate}
    \item \textbf{Better method for boson sampling}: In our simulations, GBS outperformed boson sampling in terms of the number of samples that it needed to work effectively. But since both methods use mostly similar components (beamsplitters and phaseshifters), the difference may be more to do with how we encode the problems for boson sampling.
    One naive solution could be to use the same technique to make unitary matrices that GBS uses, but by using single photon sources rather than squeezing light. However, this suggestion would probably need further refinement since (i) producing synchronized single photons at scale is a major engineering hurdle and (ii) it ignores the possibility of photon loss.
    \item \textbf{Implementation on real photonic hardware:} Another potential future work that can be done is a comparison on real photonic hardware for boson sampling and GBS. At the time of writing, there was no publicly available hardware that would support problems of the sizes like the ones in this work.
    \item \textbf{Experimentation on real-world datasets:} With better simulation software and hardware, it would be useful to test these photonic methods on datasets representing non-synthetic data that come from domains where biclustering is the most relevant.
    \item \textbf{Comparison against classical methods:} Like we mentioned in the section before, comparison against industry-standard classical methods is also part of the work that needs to be done for making stronger claims.
    \item \textbf{Develop hybrid quantum-classical techniques:} This is a direction of research that we believe will make a significant impact for this application. It is quite possible that classical methods can outperform a direct application of restricted models of quantum computing (like boson sampling and GBS), at least in the short term, when the latter has to be moderately adapted to an application problem. This is because classical methods can often exploit their Turing completeness to implement a range of solutions rather than be restricted to just one type. For example, currently, it is still challenging for quantum annealing (another restricted model of quantum computing) to solve k-SAT problems better than classical methods \cite{gabor2019assessing}.

   Another reason for investigating hybrid quantum-classical techniques is the fact that quantum circuits in the short to medium term (even for photonic-based hardware) will remain relatively small. Taking these two points together, we believe that in the best case scenario, boson sampling or GBS may be best used as a smaller (but effective) sub-routine inside a larger method for solving problems in biclustering. This is essentially to counter disadvantages of either approaches.

\end{enumerate}

\section{Conclusion}\label{sec:conclusion}
In this work, we proposed the use of two computational models from photonics : namely boson sampling and Gaussian boson sampling (GBS), for the problem of biclustering. Being the first work in this research direction, we conducted four preliminary tests where we simulated the application of these quantum computing models on synthetic datasets for biclustering. We found that these models are best suited for binary datasets and datasets where the contrast in the values is very high. We also found that the direct application of GBS has two main advantages over the direct application of boson sampling: (i) fewer number of samples needed and (ii) the ability two locate both rows and columns of a bicluster simultaneously. Based on our findings, we recommend a list of future work, primarily to do with (a) better encoding of biclustering problems in boson sampling, (b) experiments on photonic hardware and on (c) non-synthetic data, (d) comparison with classical methods and (e) the development of hybrid quantum-classical methods. We hope that the results of our preliminary simulations are useful to the research community for all future work in this domain.
\begin{credits}
\subsubsection{\ackname} The authors would like to thank Dr Charles Nicholas and the DREAM laboratory in the University of Maryland Baltimore County (UMBC), for providing access to their high performance compute servers for our simulations.

\subsubsection{\discintname}
The authors declare that the research was conducted in the absence of any commercial or financial relationships that could be construed as a potential conflict of interest.
\end{credits}
%
%
%
\bibliographystyle{splncs04}
\bibliography{references}

\begin{thebibliography}{10}
\providecommand{\url}[1]{\texttt{#1}}
\providecommand{\urlprefix}{URL }
\providecommand{\doi}[1]{https://doi.org/#1}

\bibitem{aaronson2011computational}
Aaronson, S., Arkhipov, A.: The computational complexity of linear optics. In: Proceedings of the forty-third annual ACM symposium on Theory of computing. pp. 333--342 (2011)

\bibitem{aaronson2013bosonsampling}
Aaronson, S., Arkhipov, A.: Bosonsampling is far from uniform. arXiv preprint arXiv:1309.7460  (2013)

\bibitem{adachi2015application}
Adachi, S.H., Henderson, M.P.: Application of quantum annealing to training of deep neural networks. arXiv preprint arXiv:1510.06356  (2015)

\bibitem{arrazola2018using}
Arrazola, J.M., Bromley, T.R.: Using gaussian boson sampling to find dense subgraphs. Physical review letters  \textbf{121}(3),  030503 (2018)

\bibitem{arrazola2018quantum}
Arrazola, J.M., Bromley, T.R., Rebentrost, P.: Quantum approximate optimization with gaussian boson sampling. Physical Review A  \textbf{98}(1),  012322 (2018)

\bibitem{ayadi2009biclustering}
Ayadi, W., Elloumi, M., Hao, J.K.: A biclustering algorithm based on a bicluster enumeration tree: application to dna microarray data. BioData mining  \textbf{2},  1--16 (2009)

\bibitem{bertsimas1993simulated}
Bertsimas, D., Tsitsiklis, J.: Simulated annealing. Statistical science  \textbf{8}(1),  10--15 (1993)

\bibitem{bonaldi2023boost}
Bonaldi, N., Rossi, M., Mattioli, D., Grapulin, M., Fern{\'a}ndez, B.S., Caputo, D., Magagnini, M., Osti, A., Veronese, F.: Boost clustering with gaussian boson sampling: a full quantum approach. arXiv preprint arXiv:2307.13348  (2023)

\bibitem{bottarelli2018biclustering}
Bottarelli, L., Bicego, M., Denitto, M., Di~Pierro, A., Farinelli, A., Mengoni, R.: Biclustering with a quantum annealer. Soft Computing  \textbf{22},  6247--6260 (2018)

\bibitem{brod2019photonic}
Brod, D.J., Galv{\~a}o, E.F., Crespi, A., Osellame, R., Spagnolo, N., Sciarrino, F.: Photonic implementation of boson sampling: a review. Advanced Photonics  \textbf{1}(3),  034001--034001 (2019)

\bibitem{castanho2022biclustering}
Castanho, E.N., Aidos, H., Madeira, S.C.: Biclustering fmri time series: a comparative study. BMC bioinformatics  \textbf{23}(1),  1--30 (2022)

\bibitem{de2007applying}
de~Castro, P.A., de~Fran{\c{c}}a, F.O., Ferreira, H.M., Von~Zuben, F.J.: Applying biclustering to text mining: an immune-inspired approach. In: Artificial Immune Systems: 6th International Conference, ICARIS 2007, Santos, Brazil, August 26-29, 2007. Proceedings. pp. 83--94. Springer (2007)

\bibitem{choi2018reinforcement}
Choi, S., Ha, H., Hwang, U., Kim, C., Ha, J.W., Yoon, S.: Reinforcement learning based recommender system using biclustering technique. arXiv preprint arXiv:1801.05532  (2018)

\bibitem{cipra1987introduction}
Cipra, B.A.: An introduction to the ising model. The American Mathematical Monthly  \textbf{94}(10),  937--959 (1987)

\bibitem{clements2016optimal}
Clements, W.R., Humphreys, P.C., Metcalf, B.J., Kolthammer, W.S., Walmsley, I.A.: Optimal design for universal multiport interferometers. Optica  \textbf{3}(12),  1460--1465 (2016)

\bibitem{cormen2022introduction}
Cormen, T.H., Leiserson, C.E., Rivest, R.L., Stein, C.: Introduction to algorithms. MIT press (2022)

\bibitem{deng2023solving}
Deng, Y.H., Gong, S.Q., Gu, Y.C., Zhang, Z.J., Liu, H.L., Su, H., Tang, H.Y., Xu, J.M., Jia, M.H., Chen, M.C., et~al.: Solving graph problems using gaussian boson sampling. Physical Review Letters  \textbf{130}(19),  190601 (2023)

\bibitem{deng2023gaussian}
Deng, Y.H., Gu, Y.C., Liu, H.L., Gong, S.Q., Su, H., Zhang, Z.J., Tang, H.Y., Jia, M.H., Xu, J.M., Chen, M.C., et~al.: Gaussian boson sampling with pseudo-photon-number-resolving detectors and quantum computational advantage. Physical review letters  \textbf{131}(15),  150601 (2023)

\bibitem{deshpande2022quantum}
Deshpande, A., Mehta, A., Vincent, T., Quesada, N., Hinsche, M., Ioannou, M., Madsen, L., Lavoie, J., Qi, H., Eisert, J., et~al.: Quantum computational advantage via high-dimensional gaussian boson sampling. Science advances  \textbf{8}(1),  eabi7894 (2022)

\bibitem{dhillon2001co}
Dhillon, I.S.: Co-clustering documents and words using bipartite spectral graph partitioning. In: Proceedings of the seventh ACM SIGKDD international conference on Knowledge discovery and data mining. pp. 269--274 (2001)

\bibitem{gabor2019assessing}
Gabor, T., Zielinski, S., Feld, S., Roch, C., Seidel, C., Neukart, F., Galter, I., Mauerer, W., Linnhoff-Popien, C.: Assessing solution quality of 3sat on a quantum annealing platform. In: Quantum Technology and Optimization Problems: First International Workshop, QTOP 2019, Munich, Germany, March 18, 2019, Proceedings 1. pp. 23--35. Springer (2019)

\bibitem{glynn2013permanent}
Glynn, D.G.: Permanent formulae from the veronesean. Designs, codes and cryptography  \textbf{68},  39--47 (2013)

\bibitem{halmos1951normal}
Halmos, P.R.: Normal dilations and extensions of operators. In: BULLETIN OF THE AMERICAN MATHEMATICAL SOCIETY. vol. 57, 4, pp. 294--294. AMER MATHEMATICAL SOC 201 CHARLES ST, PROVIDENCE, RI 02940-2213 (1951)

\bibitem{halmos1950summa}
Halmos, P.: Summa. brasil. math. Normal dilations and extensions of operators  \textbf{2},  125--134 (1950)

\bibitem{hamilton2017gaussian}
Hamilton, C.S., Kruse, R., Sansoni, L., Barkhofen, S., Silberhorn, C., Jex, I.: Gaussian boson sampling. Physical review letters  \textbf{119}(17),  170501 (2017)

\bibitem{hochreiter2010fabia}
Hochreiter, S., Bodenhofer, U., Heusel, M., Mayr, A., Mitterecker, A., Kasim, A., Khamiakova, T., Van~Sanden, S., Lin, D., Talloen, W., et~al.: Fabia: factor analysis for bicluster acquisition. Bioinformatics  \textbf{26}(12),  1520--1527 (2010)

\bibitem{jose2022biclustering}
Jos{\'e}-Garc{\'\i}a, A., Jacques, J., Sobanski, V., Dhaenens, C.: Biclustering algorithms based on metaheuristics: a review. Metaheuristics for Machine Learning: New Advances and Tools pp. 39--71 (2022)

\bibitem{kadowaki1998quantum}
Kadowaki, T., Nishimori, H.: Quantum annealing in the transverse ising model. Physical Review E  \textbf{58}(5), ~5355 (1998)

\bibitem{karim2019bicluso}
Karim, M.B., Huang, M., Ono, N., Kanaya, S., Altaf-Ul-Amin, M.: Bicluso: A novel biclustering approach and its application to species-voc relational data. IEEE/ACM transactions on computational biology and bioinformatics  \textbf{17}(6),  1955--1965 (2019)

\bibitem{king2019quantum}
King, J., Yarkoni, S., Raymond, J., Ozfidan, I., King, A.D., Nevisi, M.M., Hilton, J.P., McGeoch, C.C.: Quantum annealing amid local ruggedness and global frustration. Journal of the Physical Society of Japan  \textbf{88}(6),  061007 (2019)

\bibitem{kirkpatrick1983optimization}
Kirkpatrick, S., Gelatt~Jr, C.D., Vecchi, M.P.: Optimization by simulated annealing. science  \textbf{220}(4598),  671--680 (1983)

\bibitem{kluger2003spectral}
Kluger, Y., Basri, R., Chang, J.T., Gerstein, M.: Spectral biclustering of microarray data: coclustering genes and conditions. Genome research  \textbf{13}(4),  703--716 (2003)

\bibitem{kumar2018quantum}
Kumar, V., Bass, G., Tomlin, C., Dulny, J.: Quantum annealing for combinatorial clustering. Quantum Information Processing  \textbf{17},  1--14 (2018)

\bibitem{madeira2004biclustering}
Madeira, S.C., Oliveira, A.L.: Biclustering algorithms for biological data analysis: a survey. IEEE/ACM transactions on computational biology and bioinformatics  \textbf{1}(1),  24--45 (2004)

\bibitem{madsen2022quantum}
Madsen, L.S., Laudenbach, F., Askarani, M.F., Rortais, F., Vincent, T., Bulmer, J.F., Miatto, F.M., Neuhaus, L., Helt, L.G., Collins, M.J., et~al.: Quantum computational advantage with a programmable photonic processor. Nature  \textbf{606}(7912),  75--81 (2022)

\bibitem{mezher2023solving}
Mezher, R., Carvalho, A.F., Mansfield, S.: Solving graph problems with single-photons and linear optics. arXiv preprint arXiv:2301.09594  (2023)

\bibitem{miller1930history}
Miller, G.: On the history of determinants. The American Mathematical Monthly  \textbf{37}(5),  216--219 (1930)

\bibitem{mirkin1997mathematical}
Mirkin, B.: Mathematical classification and clustering. Journal of the Operational Research Society  \textbf{48}(8),  852--852 (1997)

\bibitem{orzechowski2016text}
Orzechowski, P., Boryczko, K.: Text mining with hybrid biclustering algorithms. In: International Conference on Artificial Intelligence and Soft Computing. pp. 102--113. Springer (2016)

\bibitem{pontes2015biclustering}
Pontes~Balanza, B., Gir{\'a}ldez, R., Aguilar~Ruiz, J.S.: Biclustering on expression data: A review. Journal of Biomedical Informatics, 57 (October 2015), 163-180.  (2015)

\bibitem{quesada2018gaussian}
Quesada, N., Arrazola, J.M., Killoran, N.: Gaussian boson sampling using threshold detectors. Physical Review A  \textbf{98}(6),  062322 (2018)

\bibitem{raff2020automatic}
Raff, E., Zak, R., Lopez~Munoz, G., Fleming, W., Anderson, H.S., Filar, B., Nicholas, C., Holt, J.: Automatic yara rule generation using biclustering. In: Proceedings of the 13th ACM Workshop on Artificial Intelligence and Security. pp. 71--82 (2020)

\bibitem{reck1994experimental}
Reck, M., Zeilinger, A., Bernstein, H.J., Bertani, P.: Experimental realization of any discrete unitary operator. Physical review letters  \textbf{73}(1), ~58 (1994)

\bibitem{ryser1963combinatorial}
Ryser, H.J.: Combinatorial mathematics, vol.~14. American Mathematical Soc. (1963)

\bibitem{schuld2020measuring}
Schuld, M., Br{\'a}dler, K., Israel, R., Su, D., Gupt, B.: Measuring the similarity of graphs with a gaussian boson sampler. Physical Review A  \textbf{101}(3),  032314 (2020)

\bibitem{solomons2023effect}
Solomons, N.R., Thomas, O.F., McCutcheon, D.P.: Effect of photonic errors on quantum enhanced dense-subgraph finding. Physical Review Applied  \textbf{20}(5),  054043 (2023)

\bibitem{sun2022recommendation}
Sun, J., Zhang, Y.: Recommendation system with biclustering. Big Data Mining and Analytics  \textbf{5}(4),  282--293 (2022)

\bibitem{takagi1924algebraic}
Takagi, T.: On an algebraic problem reluted to an analytic theorem of carath{\'e}odory and fej{\'e}r and on an allied theorem of landau. In: Japanese journal of mathematics: transactions and abstracts. vol.~1, pp. 83--93. The Mathematical Society of Japan (1924)

\bibitem{Termini}
Termini, S.: Imagination and Rigor: Their Interaction Along the Way to Measuring Fuzziness and Doing Other Strange Things. Springer-Verlag (2006). \doi{10.1007/88-470-0472-1_13}, \url{http://dx.doi.org/10.1007/88-470-0472-1_13}

\bibitem{troyansky1996quantum}
Troyansky, L., Tishby, N.: On the quantum evaluation of the determinant and the permanent of a matrix. Proc. Phys. Comput p.~96 (1996)

\bibitem{xie2019time}
Xie, J., Ma, A., Fennell, A., Ma, Q., Zhao, J.: It is time to apply biclustering: a comprehensive review of biclustering applications in biological and biomedical data. Briefings in bioinformatics  \textbf{20}(4),  1450--1465 (2019)

\end{thebibliography}
%




\end{document}